\documentclass[final]{ustcstep}

\usepackage[dvips]{graphicx}
\usepackage{verbatim}
\usepackage{multirow}
\usepackage{url}
\usepackage{color}
\usepackage{ulem}
\usepackage{natbib}
\usepackage[displaymath]{lineno}

\def\gsim{\;\lower4pt\hbox{${\buildrel\displaystyle >\over\sim}$}\;}
\def\lsim{\;\lower4pt\hbox{${\buildrel\displaystyle <\over\sim}$}\;}
\def\grls{\;\lower4pt\hbox{${\buildrel\displaystyle >\over <}$}\;}

\newcommand{\del}[1]{{\iffalse #1 \fi}}
\newcommand{\ve}[1]{\mathbf{#1}}
\newcommand\addr[2]{{\footnotesize \it $^{#1}$#2}\\}

\begin{document}

\title{Could the collision of CMEs in the heliosphere be super-elastic? --- Validation through three-dimensional simulations}

\author{Fang Shen,$^1$ Chenglong Shen,$^2$ Yuming Wang,$^{2,*}$ Xueshang Feng $^1$ and Changqing Xiang $^1$\\[1pt]
\addr{1}{SIGMA Weather Group, State Key Laboratory of Space Weather, Center for Space Science and Applied Research, Chinese Academy}
\addr{ }{ of Sciences, Beijing 100190, China}
\addr{2}{CAS Key Laboratory of Geospace Environment, Department of Geophysics and Planetary
Sciences, University of Science and}
\addr{ }{ Technology of China, Hefei, Anhui 230026, China} 
\addr{*}{Corresponding Author, Contact: ymwang@ustc.edu.cn}}

\maketitle
\tableofcontents

\begin{abstract}
Though coronal mass ejections (CMEs) are magnetized fully-ionized gases, 
a recent observational study of a CME collision event in 2008 November has suggested that their behavior in the 
heliosphere is like elastic balls, and their collision is 
probably super-elastic \citep{Shen_etal_2012}. 
If this is true, this finding has an obvious impact on the space weather 
forecasting because the direction and veliocity of CMEs may change. 
To verify it, we numerically study the event through three-dimensional MHD 
simulations. The nature of CMEs' collision is examined
by comparing two cases. In one case the two CMEs collide as observed, but in 
the other, they do not. Results show that the collision leads to extra kinetic energy gain by 3\%--4\% 
of the initial kinetic energy of the two CMEs. It firmly proves that the 
collision of CMEs could be super-elastic.
\end{abstract}

\section{Introduction}
Dynamic process of coronal mass ejections (CMEs) in the heliosphere
is key information for us to evaluate the CMEs' geo-effectiveness. 
But it becomes more complicated when successive CMEs 
interact in the heliosphere. Both observational
and numerical studies have shown that CME's shape, velocity and direction 
may change significantly through collisions and interactions 
\citep[e.g.,][]{Wang_etal_2002b, Wang_etal_2003c, Wang_etal_2005, Reiner_etal_2003, Farrugia_Berdichevsky_2004, 
Lugaz_etal_2005, Lugaz_etal_2009, Lugaz_etal_2012, Hayashi_etal_2006, Xiong_etal_2007, 
WuC_etal_2007, Liu_etal_2012, Temmer_etal_2012, ShenF_etal_2012a, Shen_etal_2012}. 

CMEs are magnetized plasmoids. 
In most cases, CMEs could be treated as an elastic ball in the heliosphere due to low reconnection rate, 
and the collision
between them was usually thought to be elastic or inelastic, through which
the total kinetic energy of colliding CMEs conserves or decreases. This classic collision picture 
was often used to analyze the momentum exchange during CME collisions 
\citep[e.g.,][]{Lugaz_etal_2009, Temmer_etal_2012}. But the picture is 
sometimes failed to explain observations. For example, the analysis of 2010 August
1 CME-CME interaction event suggested that the collision between CMEs is unlikely
to be elastic or perfectly inelastic \citep{Temmer_etal_2012}. 
A possible explanation is that the CME-driven shock if any may be involved in the
momentum transfer \citep{Lugaz_etal_2009}. Another explanation can be found in
a most recent work about the CME-CME interaction event during November 2--8, 2008 by 
\citet{Shen_etal_2012}, which for the first time revealed that the collision of CMEs 
could be super-elastic. A fundamental definition of super-elastic collision
is that the total kinetic energy of colliding system increases after the collision.
It is unexpectedly beyond the classic collision picture,
but well explains the observed track of the leading CME in that event.

If super-elastic collision does happen, CME's effect on space weather needs to be 
re-evaluated because more thermal and magnetic
energy inside CMEs will be converted into kinetic energy, which may cause the changes of the direction and 
velocity of CMEs different from usually expected. 
However, at present, the finding of super-elastic is doubtable,
because the result was obtained based on the remote imaging
data from STEREO spacecraft and 
some highly ideal assumptions. Thus a numerical simulation may favor us validating the possibility
of CMEs' super-elastic collision.

In this letter, we carry out three-dimensional (3-D) MHD simulations based on 
the observations of the 2008 November event, and try to reveal the nature 
of the CMEs' collision through the analysis of the energy transformation during the 
collision. In the next section, the MHD model and simulation method are introduced. 
The simulation results of the CMEs' collision and a comparison with a non-collision
case are presented in Section \ref{sec_result} and \ref{sec_comp}, respectively.
In the last section, a summary and discussion is given.

\begin{figure*}[th]
  \centering
  \includegraphics[width=1.\hsize]{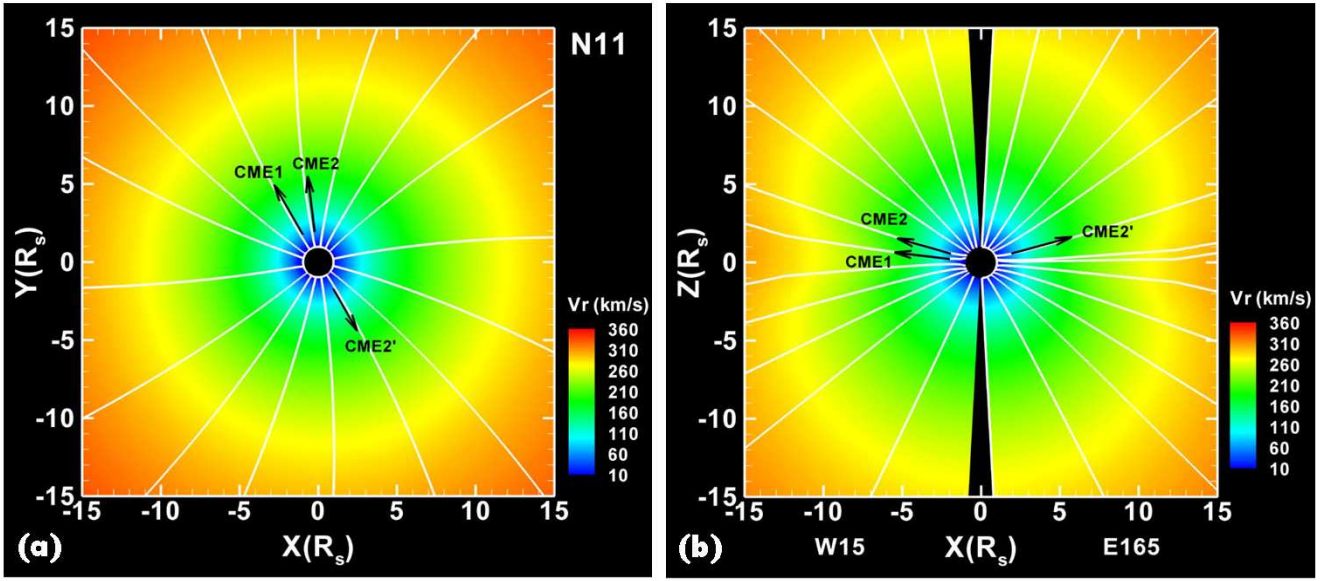}
  \caption{Background solar wind in (a) the plane of the latitude of N11$^\circ$ and 
(b) the meridian plane passing through the longitude of W15$^\circ$ and E165$^\circ$. 
The white lines show the magnetic field lines. The propagation directions of CMEs
to be introduced are indicated by arrows.}\label{fg_bk}
\end{figure*}

\section{MHD model and simulation method}
The numerical scheme we used is a 3-D corona-interplanetary total variation diminishing
(COIN-TVD) scheme in a Sun-centered spherical coordinate system $(r, \theta, \varphi)$
\citep{Feng_etal_2003, Feng_etal_2005, ShenF_etal_2007, ShenF_etal_2009}.
The projected characteristic boundary conditions \citep{Wu_Wang_1987, Hayashi_2005, Wu_etal_2006} 
are adopted at the lower boundary. 
The computational domain is set to cover $1R_s \leq r \leq 100 R_s$, 
$-89^\circ\leq\theta\leq 89^\circ$ and $0^\circ\leq\varphi\leq 360^\circ$, where $r$ is the 
radial distance from solar center in units of solar radius $R_s$, and $\theta$ and $\varphi$ are the
elevation and azimuthal angles, respectively.

We first establish a steady state of background solar wind.
The potential field extrapolated from the observed line-of-sight magnetic field 
on the photosphere and Parker's solar wind solution are used as the initial magnetic field
and velocity. The initial
density is deduced from the momentum conservation law, and the initial temperature is 
given by assuming an adiabatic process. With these initial conditions, our MHD code
may quickly reach a self-consistent and steady state of solar wind. Figure~\ref{fg_bk}
presents the radial velocity of background solar wind and magnetic field lines, which shows 
the typical characteristics, e.g., nearly axial-symmetric and dipolar, at solar minimum.

\begin{table*}[tb]
\begin{center}
\caption{Initial parameters of CMEs and background solar wind}\label{tb_cmes}
\tabcolsep 3pt
\begin{tabular}{c|ccccccc|ccccc}
\hline
& D & $v$ & $n$ & $T$ & $B$ & $\beta$ & $R$ & $E_k$ & $E_m$ & $E_i$ & $E_g$ & $E_t$\\
& & km s$^{-1}$ & $\times10^7$ cm$^{-3}$ & $\times10^5$ K & $\times10^5$ nT & &$R_s$ & \multicolumn{5}{c}{$\times10^{32}$ erg}\\
\hline
CME1 & N06W28  & 243 & 4.0 & 3.33 & 1.22 & 0.06 & 0.5 & 0.077 & 0.104 & 0.097 & $-0.064$ & 0.213 \\
CME2 & N16W08  & 407 & 5.0 & 4.17 & 1.47 & 0.06 & 0.5 & 0.261 & 0.150 & 0.145 & $-0.088$ & 0.468 \\
SW & N11W18 & $316\sim433$ & & & & & & 5.30 & 3.11 & 7.28 & $-2.52$ & 13.2 \\
\hline
\end{tabular}\\
The columns from the second one to the right are the propagation direction,
velocity, number density, temperature, magnetic field,
plasma beta, radius, and the kinetic, magnetic, thermal, gravitational and total energies, respectively.
The values of the velocity of solar wind are at $r=18$ and $100 R_s$, respectively, in the direction
of N11W18. The energies of solar wind are the integration over the whole computational domain 
before CMEs are introduced.
\end{center}
\end{table*}

As we did in the previous work, two CMEs are modeled as
magnetic blobs \citep{Chane_etal_2005, ShenF_etal_2011a}, 
and introduced successively with a separation time of 6 hours and their centers sitting at $r=2 R_s$.
Hereafter, we use CME1 and CME2 for the first and second 
initiated CMEs. To reproduce the 2008 November event, two key parameters, 
their initial propagation directions and velocities, are chosen to be the same as those derived from observations \citep{Shen_etal_2012}. 
The directions of the two CMEs are N06W28 and N16W08, respectively, and the propagation speeds 
are 243 and 407 km s$^{-1}$, respectively. Another important parameter, plasma beta, is set a reasonable
value of 0.06 for both CMEs. According to the analysis of errors in \citet{Shen_etal_2012}, 
other parameters are not pivotal, and therefore set arbitrarily.
Table \ref{tb_cmes} lists the initial parameters of the two CMEs.

\section{Simulation results}\label{sec_result}
\begin{figure*}[p]
  \centering
  \includegraphics[width=0.9\hsize]{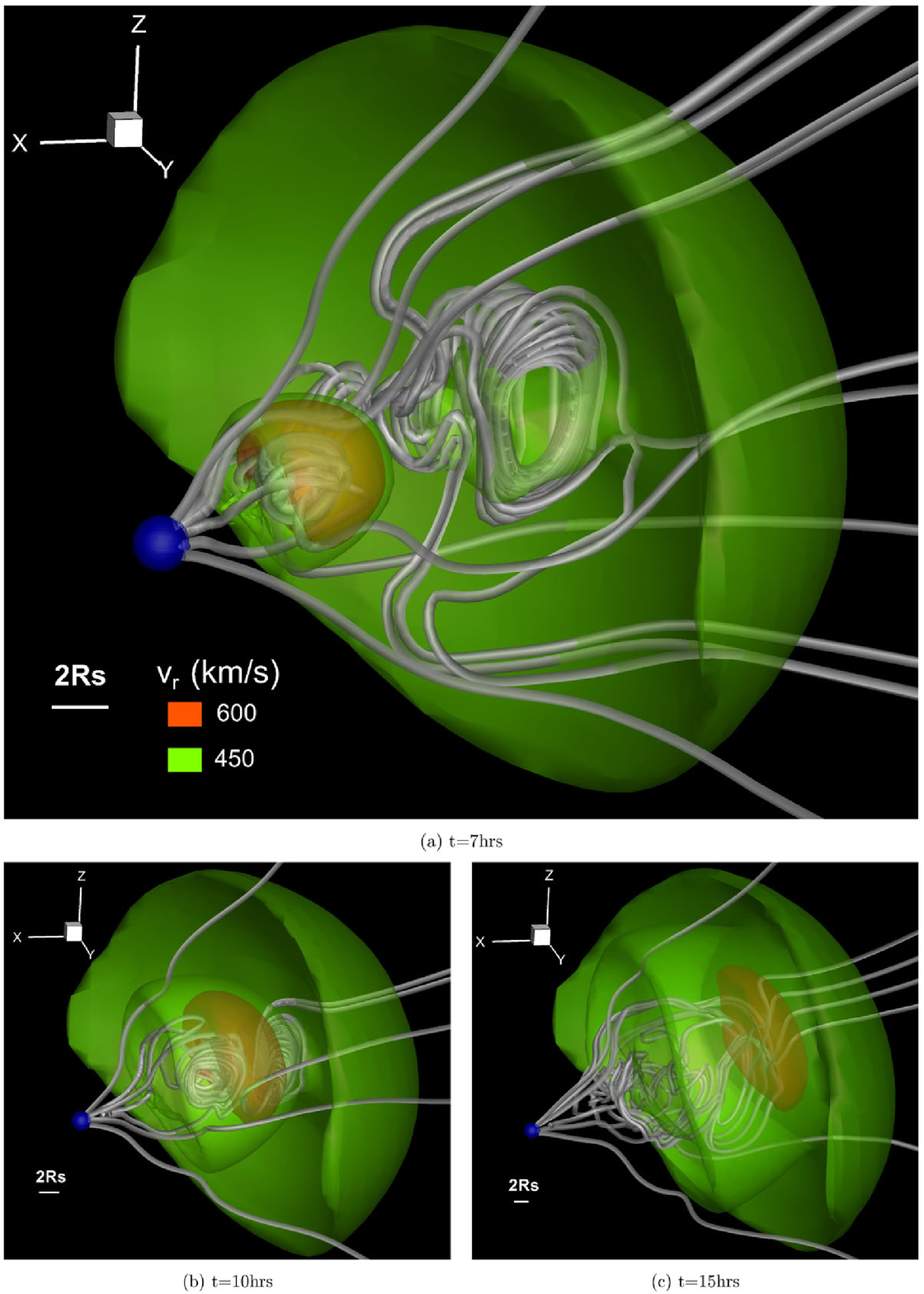}
  \caption{Radial velocity map of the two CMEs at the time of 
7, 10 and 15 hours. The surfaces of the radial velocity being 450 
and 600 km s$^{-1}$ are displayed by different colors. Some magnetic
field lines are shown as the thick white lines. The small blue ball 
shows the position and size of the Sun.}\label{fg_3dsim}
\end{figure*}

The background solar wind between the directions of the two CMEs gradually increases from about 
316 km s$^{-1}$ at $18 R_s$ to about 433 km s$^{-1}$ at $100 R_s$. Due to the expansion, 
the leading edges of the two CMEs move faster than ambient solar wind. Thus we
locate the CMEs by simply setting a threshold of 450 km s$^{-1}$ in the map of radial 
velocity. The time of introducing CME1 into computational domain is set to be zero. 
Figure~\ref{fg_3dsim} shows the 3-D view of the radial velocity distribution at 
$t=$7, 10 and 15 hours, respectively. Only the regions of 
the radial velocity equal to 450 and 600 km s$^{-1}$ are displayed for clarity. Due to
the selection effect, some shell structures are shown, but they do not reflect the real 
CME shape. The CMEs can be recognized through the superimposed node-shaped magnetic field 
lines. 

Since CME2 is faster than CME1, the two CMEs get closer and closer as shown
in the three panels. The momentum transfer could be clearly seen by noting the orange 
region. At 7 hours, right before the collision, the orange region, 
which denotes a radial velocity
of 600 km s$^{-1}$, locates in CME2. After the two CMEs touch, the orange region moves forward,
which suggests a momentum transfer from CME2 to CME1.

With some limits of the MHD code, however, we cannot identify the exact boundary of a 
CME. Thus we do not analyze the momentum or energy change for individual CMEs, but instead,
analyze the variations of all kinds of energies integrated over the whole computational 
domain. All the energies of the two CMEs and solar wind at initial time are shown
in Table~\ref{tb_cmes}. Although the energy of the two CMEs is only about 5\% of the total
energy of background solar wind, it is larger than the
errors unavoidably from numerical calculations and ideal MHD assumptions as will be seen below. 


\begin{figure}[tb]
  \centering
  \includegraphics[width=0.9\hsize]{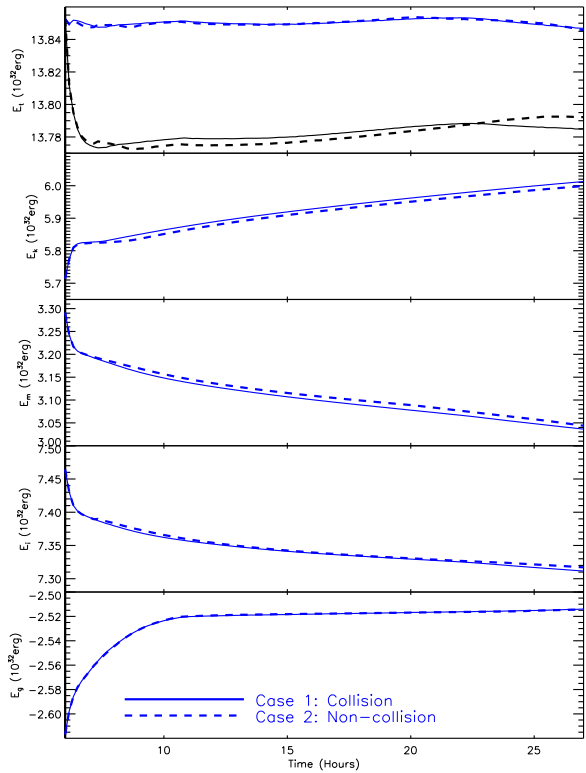}
  \caption{Temporal profiles of all kinds of energies. The panels from the top
to bottom show the total energy $E_t$, kinetic energy $E_k$, magnetic energy $E_m$,
thermal energy $E_i$ and gravitational energy $E_g$, respectively.
In the top panel, the black lines shows the total energy before correction 
(see main text for details).}\label{fg_energy}
\end{figure}

The solid black line in the upper panel of Figure~\ref{fg_energy} shows the variation of the 
total energy, $E_t$, an integrated value over the whole computational domain, after the 
launch of CME2 at $t=6$ hours. The quick drop of $E_t$ at the beginning is because the introduced CME
expels the ambient solar wind. This is a numerical effect and brings 
difficulty into the analysis of energy variation. To reduce it, we first calculate
the net energy flowing into the computational domain at boundaries in a time interval $\Delta t$,
which is
$E_b=\Delta t\int \varepsilon_t\rho\ve{v}\cdot d\ve{S}$,
where $\varepsilon_t$ is the energy density at time $t$ and $\ve{S}$ is the surface of the boundaries,
and then deduct it from the total energy to get a corrected energy. Assume that the total energy at 
any given instant $t_i$ is $E_{ti}$ and the net energy flow across the boundaries since 
the last instant $t_{i-1}$ is $E_{bi}$, the correct
total energy is 
$E_t=E_{ti}-\Sigma_1^iE_{bi}$,
which should be always equal to the total energy at initial time $t_0$ in theory. After the 
correction, the total energy varies in small range of about $5\times10^{29}$ erg as shown by 
the solid blue line in the upper panel of Figure~\ref{fg_energy}, that just indicates the numerical
error in our simulation. It is much smaller than the CME energies listed in Table~\ref{tb_cmes}. 

All kinds of energies after the correction are shown in the other panels in Figure~\ref{fg_energy}.
After the two CMEs propagate into the computational domain, the kinetic energy, $E_k$, 
and gravitational energy, $E_g$, both continuously increase, whereas the magnetic energy, $E_m$, and 
thermal energy, $E_i$, both decrease. The changes of these energies are all one order larger 
than the variation of total energy, suggesting a real physical process. The increase of $E_g$ 
is due to that the CMEs carry a heavier plasma than background solar wind. The changes of other 
energies are consistent with the well-known picture that CME's magnetic and thermal energy will 
be converted into kinetic energy as it expands during the propagation 
\citep[e.g.,][]{Kumar_Rust_1996, Wang_etal_2009}.

In order to validate that the kinetic energy gain (or partial of it) comes from a super-elastic 
collision, we need another case for comparison, in which the two CMEs do not collide. To do 
this, we adjust the longitude of CME2 to $E165^\circ$, which causes the longitudinal separation 
between the two CMEs to be $175^\circ$, and keep all the other parameters exactly the same as those in
the case of collision. Hereafter we use Case 1 for collision, Case 2 for non-collision and CME2$'$ for 
the second CME in Case 2. Figure~\ref{fg_bk} has shown that the background solar wind and magnetic structure
around CME2 and CME2$'$ are quite similar. We believe that the two cases are comparable.

\section{Comparison between the cases of collision and non-collision}\label{sec_comp}
\begin{figure}[tb]
  \centering
  \includegraphics[width=\hsize]{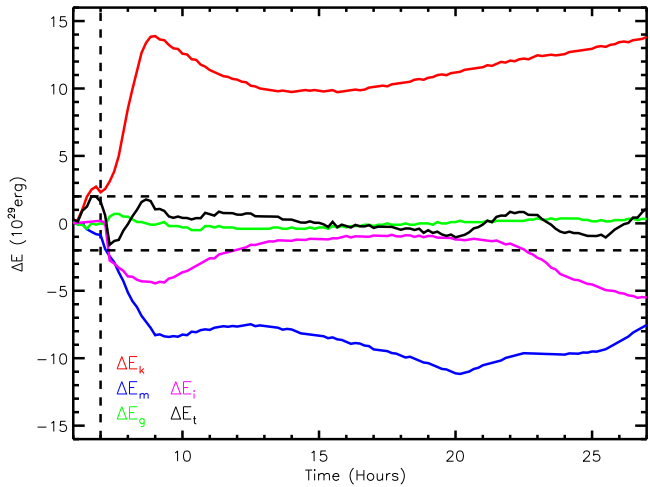}
  \caption{Energy difference between the case of collision (Case 1) and the 
case of non-collision (Case 2). A positive value means that the energy in 
Case 1 is larger than that in Case 2. The vertical dashed line marks the
beginning of the collision, and the horizontal dashed lines indicate the level of numerical error.}\label{fg_diff}
\end{figure}

From CME1 being introduced into computational domain to the instance of CME2 being 
introduced, the two cases are exactly the same. 
After CME2 is introduced, the two cases become different. The dashed blue lines in 
Figure~\ref{fg_energy} show the energy variations for Case 2, which are similar to
those in Case 1 except some small differences. These small differences are shown much clearly 
in Figure~\ref{fg_diff}.

The difference of the total energy, $\Delta E_t$, between the two cases has small 
fluctuations with an amplitude of about $2\times10^{29}$ erg. It indicates the 
level of numerical error. The difference of the gravitational
energy, $\Delta E_g$, is about $10^{29}$ erg, smaller than the numerical error. Thus we cannot 
conclude if $\Delta E_g$ is real or not. For all the other energies, the differences are 
significantly larger than the error, and thought to be physically meaningful. 

It is found that, from the time of $t=7$ hours, the difference of the kinetic energy, $\Delta E_k$,
rapidly increases from about $2\times10^{29}$ erg to about $1.4\times10^{30}$ erg in 2 hours, 
and then decreases back to about $10^{30}$ erg and slowly returns. It means that
there is extra kinetic energy gain in Case 1. Recall that the energy flow across
the boundaries has been deducted, and therefore the extra kinetic energy gain must come
from the collision of the two CMEs. Although we do not know the kinetic energy for each CME, the comparison
between Case 2 and Case 1 is just like the comparison between the state before and after 
the collision. The significant difference between the two cases in the kinetic energy
does confirm that the collision of CMEs could be super-elastic as suggested by \citet{Shen_etal_2012}.

It is hard to identify when the collision ends. It might be at $t=20$ hours or even later.
But we are sure that the two CMEs have fully interacted for a long time. This long process 
allows magnetic and thermal energies to be converted into kinetic energy.
It is noticed that the decrease of the magnetic energy is much larger 
than that of the thermal energy, which suggests that the magnetic energy stored in CMEs
are the major source of the extra kinetic energy gain.

\section{Summary and discussion}
We have comparatively investigated the energy variation during the collision of two
successive CMEs. It is found that the kinetic energy gain in 
the case of collision is larger than that in the case of non-collision though the initial conditions
of the two CMEs and the background solar wind are exactly the same.
This result does
suggest that the collision between the two CMEs is super-elastic, through which additional
magnetic and thermal energies are converted into kinetic energy. 

In this study, the initial kinetic energy of the two CMEs is about $33.8\times10^{30}$ erg 
(see Table~\ref{tb_cmes}). Since the collision happens quickly after the introductions of the CMEs,
we may use this value approximately as the CMEs' kinetic energy right before the collision. 
The extra kinetic energy gain due to the collision is on the order of $10^{30}$ erg. It is
therefore derived that the super-elastic collision of the two CMEs causes their total kinetic 
energy increased by about 3\%--4\%, which is close to the value of 6.6\% given by \citet{Shen_etal_2012}.
Assuming the energy gain totally goes to CME1, we then estimate that the kinetic energy 
of CME1 increases by about 13\%. Normally, the leading CME will be accelerated and the trailing 
CME decelerated \citep[e.g.,][]{Wang_etal_2005, ShenF_etal_2012a, Lugaz_etal_2012}. 
Thus the percentage of the kinetic energy gain of CME1 should be even higher.
In terms of velocity, CME1 is speeded up by at least 6\%, i.e., 15 km s$^{-1}$. This number
is not large enough to impact the space weather forecasting. But a comprehensive investigation of 
the effect of collision on the velocity and direction of CMEs is still worth being pursued.

In this letter, we only consider the CMEs
similar to the 2008 November event. It is not clear if the collision between any CMEs is super-elastic. 
Moreover, some open questions remain. For example, how are the magnetic or thermal energies convert 
into kinetic energy? How does magnetic reconnection influence
the collision process and result if it efficiently occurred?
Another interesting thing is that the 2010 August event studied by \citet{Temmer_etal_2012} might be a case of `super-inelastic'
collision, a process somewhat like merging, of two fast CMEs. How and why could it happen?
All these questions are worthy of further studies.

\acknowledgments{
This work is jointly supported by grants from the 973 key projects (2012CB825601, 2011CB811403), 
the CAS Knowledge Innovation Program (KZZD-EW-01-4), the NFSC (41031066, 41074121, 41231068, 
41174150, 41274192, 41131065, 41121003 and 41274173), the Specialized Research Fund for State Key 
Laboratories, and the Public science and technology research funds projects of ocean (201005017).
}


\begin{thebibliography}{29}
\providecommand{\natexlab}[1]{#1}
\expandafter\ifx\csname urlstyle\endcsname\relax
  \providecommand{\doi}[1]{doi:\discretionary{}{}{}#1}\else
  \providecommand{\doi}{doi:\discretionary{}{}{}\begingroup
  \urlstyle{rm}\Url}\fi

\bibitem[{\textit{Chan\'e et~al.}(2005)\textit{Chan\'e, Jacobs, {Van der
  Holst}, Poedts, and Kimpe}}]{Chane_etal_2005}
Chan\'e, E., C.~Jacobs, B.~{Van der Holst}, S.~Poedts, and D.~Kimpe, On the
  effect of the initial magnetic polarity and of the background wind on the
  evolution of {CME} shocks, \textit{Astron. \& Astrophys.}, \textit{432},
  331--339, 2005.

\bibitem[{\textit{Farrugia and
  Berdichevsky}(2004)}]{Farrugia_Berdichevsky_2004}
Farrugia, C., and D.~Berdichevsky, Evolutionary signatures in complex ejecta
  and their driven shocks, \textit{Ann. Geophys.}, \textit{22}, 3679--3698,
  2004.

\bibitem[{\textit{Feng et~al.}(2003)\textit{Feng, Wu, Wei, and
  Fan}}]{Feng_etal_2003}
Feng, X., S.~T. Wu, F.~Wei, and Q.~Fan, A class of {TVD} type combined
  numerical scheme for {MHD} equations with a survey about numerical methods in
  solar wind simulations, \textit{Space Sci. Rev.}, \textit{107}, 43--53, 2003.

\bibitem[{\textit{Feng et~al.}(2005)\textit{Feng, Xiang, Zhong, and
  Fan}}]{Feng_etal_2005}
Feng, X., C.~Xiang, D.~Zhong, and Q.~Fan, A comparative study on 3-d solar wind
  structure observed by {Ulysses} and {MHD} simulation, \textit{Chinese Sci.
  Bull.}, \textit{50}, 672--678, 2005.

\bibitem[{\textit{Hayashi}(2005)}]{Hayashi_2005}
Hayashi, K., Magnetohydrodynamic simulations of the solar corona and solar wind
  using a boundary treatment to limit solar wind mass flux, \textit{Astrophys.
  J.}, \textit{161}, 480--494, 2005.

\bibitem[{\textit{Hayashi et~al.}(2006)\textit{Hayashi, Zhao, and
  Liu}}]{Hayashi_etal_2006}
Hayashi, K., X.-P. Zhao, and Y.~Liu, {MHD} simulation of two successive
  interplanetary disturbances driven by cone-model parameters in {IPS}-based
  solar wind, \textit{Geophys. Res. Lett.}, \textit{33}, L20,103, 2006.

\bibitem[{\textit{Kumar and Rust}(1996)}]{Kumar_Rust_1996}
Kumar, A., and D.~M. Rust, Interplanetary magnetic clouds, helicity
  conservation, and current-core flux-ropes, \textit{J. Geophys. Res.},
  \textit{101}, 15,667--15,684, 1996.

\bibitem[{\textit{Liu et~al.}(2012)\textit{Liu, Luhmann, M{\"{o}}stl,
  {Mart{\'{\i}}nez-Oliveros}, Bale, Lin, Harrison, Temmer, Webb, and
  Odstrcil}}]{Liu_etal_2012}
Liu, Y.~D., J.~G. Luhmann, C.~M{\"{o}}stl, J.~C. {Mart{\'{\i}}nez-Oliveros},
  S.~D. Bale, R.~P. Lin, R.~A. Harrison, M.~Temmer, D.~F. Webb, and
  D.~Odstrcil, Interactions between coronal mass ejections viewed in
  coordinated imaging and in situ observations, \textit{Astrophys. J.},
  \textit{746}, L15, 2012.

\bibitem[{\textit{Lugaz et~al.}(2005)\textit{Lugaz, Manchester, and
  Gombosi}}]{Lugaz_etal_2005}
Lugaz, N., I.~Manchester, W.~B., and T.~I. Gombosi, Numerical simulation of the
  interaction of two coronal mass ejections from {Sun} to {Earth},
  \textit{Astrophys. J.}, \textit{634}, 651--662, 2005.

\bibitem[{\textit{Lugaz et~al.}(2009)\textit{Lugaz, Vourlidas, and
  Roussev}}]{Lugaz_etal_2009}
Lugaz, N., A.~Vourlidas, and I.~I. Roussev, Deriving the radial distances of
  wide coronal mass ejections from elongation measurements in the heliosphere
  application to {CME-CME} interaction, \textit{Ann. Geophys.}, \textit{27},
  3479--3488, 2009.

\bibitem[{\textit{Lugaz et~al.}(2012)\textit{Lugaz, Farrugia, Davies, Mostl,
  Davis, Roussev, and Temmer}}]{Lugaz_etal_2012}
Lugaz, N., C.~J. Farrugia, J.~A. Davies, C.~Mostl, C.~J. Davis, I.~I. Roussev,
  and M.~Temmer, The deflection of the two interacting coronal mass ejections
  of 2010 may 23­24 as revealed by combined in situ measurements and
  heliospheric imaging, \textit{Astrophys. J.}, \textit{759}, 68(13pp), 2012.

\bibitem[{\textit{Reiner et~al.}(2003)\textit{Reiner, Vourlidas, {St. Cyr},
  Burkepile, Howard, Kaiser, Prestage, and Bougeret}}]{Reiner_etal_2003}
Reiner, M.~J., A.~Vourlidas, O.~C. {St. Cyr}, J.~T. Burkepile, R.~A. Howard,
  M.~L. Kaiser, N.~P. Prestage, and J.-L. Bougeret, Constraints on coronal mass
  ejection dynamics from simultaneous radio and white-light observations,
  \textit{Astrophys. J.}, \textit{590}, 533--546, 2003.

\bibitem[{\textit{C. Shen et~al.}(2012)\textit{Shen, Wang, Wang,
  Liu, Liu, Vourlidas, Miao, Ye, Liu, and Zhou}}]{Shen_etal_2012}
Shen, C., Y.~Wang, S.~Wang, Y.~Liu, R.~Liu, A.~Vourlidas, B.~Miao, P.~Ye,
  J.~Liu, and Z.~Zhou, Super-elastic collision of large-scale magnetized
  plasmoids in the heliosphere, \textit{Nature Phys.}, \textit{8}, 923--928,
  2012.

\bibitem[{\textit{Shen et~al.}(2007)\textit{Shen, Feng, Wu, and
  Xiang}}]{ShenF_etal_2007}
Shen, F., X.~Feng, S.~T. Wu, and C.~Xiang, Three-dimensional {MHD} simulation
  of {CMEs} in three-dimensional background solar wind with the self-consistent
  structure on the source surface as input: Numerical simulation of the
  {January} 1997 {Sun-Earth} connection event, \textit{J. Geophys. Res.},
  \textit{112}, A06,109, 2007.

\bibitem[{\textit{Shen et~al.}(2009)\textit{Shen, Feng, and
  Song}}]{ShenF_etal_2009}
Shen, F., X.~Feng, and W.~B. Song, An asynchronous and parallel time-marching
  method: application to the three-dimensional {MHD} simulation of the solar
  wind, \textit{Science in china Series E: Technological Sciences},
  \textit{52}, 2895--2902, 2009.

\bibitem[{\textit{Shen et~al.}(2011)\textit{Shen, Feng, Wang, Wu,
  Song, Guo, and Zhou}}]{ShenF_etal_2011a}
Shen, F., X.~S. Feng, Y.~Wang, S.~T. Wu, W.~B. Song, J.~P. Guo, and Y.~F. Zhou,
  Three-dimensional {MHD} simulation of two coronal mass ejections' propagation
  and interaction using a successive magnetized plasma blobs model, \textit{J.
  Geophys. Res.}, \textit{116}, A09,103, 2011.

\bibitem[{\textit{Shen et~al.}(2012)\textit{Shen, Wu, Feng, and
  Wu}}]{ShenF_etal_2012a}
Shen, F., S.~T. Wu, X.~Feng, and C.-C. Wu, Acceleration and deceleration of
  coronal mass ejections during propagation and interaction, \textit{J.
  Geophys. Res.}, \textit{117}, A11,101, 2012.

\bibitem[{\textit{Temmer et~al.}(2012)\textit{Temmer, Vr\v{s}nak, Rollett,
  Bein, {de Koning}, Liu, Bosman, Davies, M{\"{o}}stl, \v{Z}ic, Veronig,
  Bothmer, Harrison, Nitta, Bisi, Flor, Eastwood, Odstrcil, and
  Forsyth}}]{Temmer_etal_2012}
Temmer, M., B.~Vr\v{s}nak, T.~Rollett, B.~Bein, C.~A. {de Koning}, Y.~Liu,
  E.~Bosman, J.~A. Davies, C.~M{\"{o}}stl, T.~\v{Z}ic, A.~M. Veronig,
  V.~Bothmer, R.~Harrison, N.~Nitta, M.~Bisi, O.~Flor, J.~Eastwood,
  D.~Odstrcil, and R.~Forsyth, Characteristics of kinematics of a coronal mass
  ejection during the 2010 {August} 1 {CME-CME} interaction event,
  \textit{Astrophys. J.}, \textit{749}, 57, 2012.

\bibitem[{\textit{Wang et~al.}(2005)\textit{Wang, Zheng, Wang, and
  Ye}}]{Wang_etal_2005}
Wang, Y., H.~Zheng, S.~Wang, and P.~Ye, {MHD} simulation of the formation and
  propagation of multiple magnetic clouds in the heliosphere, \textit{Astron.
  \& Astrophys.}, \textit{434}, 309--316, 2005.

\bibitem[{\textit{Wang et~al.}(2009)\textit{Wang, Zhang, and
  Shen}}]{Wang_etal_2009}
Wang, Y., J.~Zhang, and C.~Shen, An analytical model probing the internal state
  of coronal mass ejections based on observations of their expansions and
  propagations, \textit{J. Geophys. Res.}, \textit{114}, A10,104, 2009.

\bibitem[{\textit{Wang et~al.}(2002)\textit{Wang, Wang, and
  Ye}}]{Wang_etal_2002b}
Wang, Y.~M., S.~Wang, and P.~Z. Ye, Multiple magnetic clouds in interplanetary
  space, \textit{Sol. Phys.}, \textit{211}, 333--344, 2002.

\bibitem[{\textit{Wang et~al.}(2003)\textit{Wang, Ye, and
  Wang}}]{Wang_etal_2003c}
Wang, Y.~M., P.~Z. Ye, and S.~Wang, Multiple magnetic clouds: Several examples
  during {March} -- {April}, 2001, \textit{J. Geophys. Res.}, \textit{108},
  1370, 2003.

\bibitem[{\textit{Wu et~al.}(2007)\textit{Wu, Fry, Dryer, Wu, Thompson, Liou,
  and Feng}}]{WuC_etal_2007}
Wu, C.-C., C.~D. Fry, M.~Dryer, S.~T. Wu, B.~Thompson, K.~Liou, and X.~S. Feng,
  Three-dimensional global simulation of multiple {ICMEs} interaction and
  propagation from the {Sun} to the heliosphere following the 25–28 {October}
  2003 solar events, \textit{Adv. in Space Res.}, \textit{40}, 1827--1834,
  2007.

\bibitem[{\textit{Wu and Wang}(1987)}]{Wu_Wang_1987}
Wu, S.~T., and J.~F. Wang, Numerical tests of a modified full implicit eulerian
  scheme with projected normal characteristic boundary conditions for {MHD}
  flows, \textit{Comput. Methods Appl. Mech. Eng.}, \textit{24}, 267--282,
  1987.

\bibitem[{\textit{Wu et~al.}(2006)\textit{Wu, Wang, Liu, and
  Hoeksema}}]{Wu_etal_2006}
Wu, S.~T., A.~H. Wang, Y.~Liu, and J.~T. Hoeksema, Data driven
  magnetohydrodynamic model for active region evolution, \textit{Astrophys.
  J.}, \textit{652}, 800--811, 2006.

\bibitem[{\textit{Xiong et~al.}(2007)\textit{Xiong, Zheng, Wu, Wang, and
  Wang}}]{Xiong_etal_2007}
Xiong, M., H.~Zheng, S.~T. Wu, Y.~Wang, and S.~Wang, Magnetohydrodynamic
  simulation of the interaction between two interplanetary magnetic clouds and
  its consequent geoeffectiveness, \textit{J. Geophys. Res.}, \textit{112},
  A11,103, 2007.

\end{thebibliography}
\end{document}